\documentclass[twocolumn]{emulateapj}

\usepackage{graphicx}
\usepackage{natbib}
\usepackage{multirow}
\usepackage{textcomp}
\usepackage{longtable}
\usepackage{url}

\newcommand{\kms}{km~s$^{-1}$}
\newcommand{\am}{NH$_{3}$}
\newcommand{\cyano}{HC$_3$N}

\newcommand{\meth}{CH$_3$OH}
\newcommand{\methts}{CH$_3$OH (4$_{-1}$--3$_0$)}
\newcommand{\methff}{CH$_3$OH (7$_{0}$--6$_1$)}
\newcommand{\methtst}{(4$_{-1}$--3$_0$)}
\newcommand{\methfft}{(7$_{0}$--6$_1$)}
\newcommand{\methcy}{CH$_3$CN}

\newcommand{\form}{H$_2$CO}
\newcommand{\jyb}{Jy beam$^{-1}$}
\newcommand{\mjyb}{mJy beam$^{-1}$}
\newcommand{\mjybk}{mJy beam$^{-1}$ km s$^{-1}$}
\newcommand{\hii}{{H \sc{II}}}
\begin{document}

\slugcomment{}

\title{The Unusual Galactic Center Radio Source N3}

\author{D.A. Ludovici}
\affil{Department of Physics and Astronomy, University of Iowa, Iowa City, IA 52245}
\email{dominic-ludovici@uiowa.edu}

\author{C.C. Lang}
\affil{Department of Physics and Astronomy, University of Iowa, Iowa City, IA 52245}

\author{M.R. Morris}
\affil{Department of Physics and Astronomy, University of California, Box 951547, 430 Portola Plaza, Los Angeles, CA 90095-1547}

\author{R. Mutel}
\affil{Department of Physics and Astronomy, University of Iowa, Iowa City, IA 52245}

\author{E.A.C. Mills}
\affil{National Radio Astronomy Observatory\altaffilmark{1} 1003 Lopezville Rd Socorro, NM 87801}

\author{J.E. Toomey IV}
\affil{Department of Science, United States Coast Guard Academy, 27 Mohegan Ave New London, CT, 06320 } 

\author{J. Ott}
\affil{National Radio Astronomy Observatory\altaffilmark{1} 1003 Lopezville Rd Socorro, NM 87801}

\altaffiltext{1}{The National Radio Astronomy Observatory is a facility of the National Science Foundation operated under cooperative agreement by Associated Universities, Inc.}

\begin{abstract}

Here we report on new, multi-wavelength radio observations of the unusual point source ``N3'' that appears to be located in the vicinity of the Galactic Center.  VLA observations between 2 and 50 GHz reveal that N3 is a compact and bright source (56 mJy at 10 GHz) with a non-thermal spectrum superimposed upon the non-thermal radio filaments (NTFs) of the Radio Arc. Our highest frequency observations place a strict upper limit of 65$\times$28 mas on the size of N3. We compare our observations to those of \citet{Yusef-Zadeh1987} and \citet{Lang1997} and conclude that N3 is variable over long time scales. Additionally, we present the detection of a compact molecular cloud located adjacent to N3 in projection. \methcy{}, \meth{}, CS, \cyano{}, HNCO, SiO, SO, and \am{} are detected in the cloud and most transitions have FWHM line widths of $\sim 20$ \kms{}. The rotational temperature determined from the metastable \am{} transitions ranges from 79 K to 183 K depending on the transitions used. We present evidence that this molecular cloud is interacting with N3. After exploring the relationship between the NTFs, molecular cloud, and N3, we conclude that N3 likely lies within the Galactic Center. We are able to rule out the \hii{} region, young supernova, active star, AGN, and micro-quasar hypotheses for N3. While a micro-blazar may provide a viable explanation for N3, additional observations are needed to determine the physical counterpart of this mysterious source. 

\end{abstract}


\section{Introduction}

The center of our Galaxy hosts a number of unique features not observed elsewhere in the Galactic disk. One of the most unusual regions in the Galactic Center (GC) is the Radio Arc region, lying $\sim$30 pc in projection from the dynamical center of the Galaxy, coincident with the radio counterpart of the Galactic black hole, SgrA$^*$. The Radio Arc contains both thermal sources (\hii{} regions, molecular clouds) as well as a collection of long ($\sim40$ pc), narrow ($\sim0.1$ pc) non-thermal filaments (NTFs). The Radio Arc NTFs are aligned roughly perpendicular to the Galactic plane \citep{Yusef-Zadeh1984} and appear to be interacting with the Sickle and Pistol \hii{} regions \citep{Yusef-Zadeh1987,Lang1997}. 

Located (in projection) near the middle of the Radio Arc NTFs is an unusual, bright radio point source designated as N3. First observed by \citet{Yusef-Zadeh1987} at 4.9 GHz, N3 appears to be located in the brightest filament of the Radio Arc NTFs and is well separated from the Sickle and Pistol \hii{} regions. With the exception of the NTFs, no extended continuum emission is observed immediately surrounding N3.  Figure 1 shows a finding chart of the various sources in this region. Despite the intriguing location of N3, the source has received little attention since it was first observed. Several radio studies of the GC in subsequent years have detected N3, but have not focused on the nature of this source  \citep{Lang1997,LaRosa2000,Yusef-Zadeh2004}.  \citet{Yusef-Zadeh1987} consider that N3 could be a background source, except for its suggestive position on the brightest NTF and a faint ``wake'' of emission lying between N3 and the northern half of the Sickle H{\scriptsize II} region (see Figure 1). 
More recently, H$^{13}$CO$^+$ and SiO emission lines were surveyed in this region by \citet{Tsuboi2011}, who found a compact SiO emission component near the location of N3.  In this component the brightness temperature ratio, $R_{\text{SiO/H}^{13}\text{CO}^+}$, is high ($\sim 4$), indicating that the molecular gas is shocked. \citet{Tsuboi2011} conclude that N3 might be interacting physically with the molecular gas.

Using the Karl J. Jansky Very Large Array (VLA) of the National Radio Astronomy Observatory,\footnotemark[1]
 we have conducted a large-scale study of the Radio Arc NTFs and surrounding regions. These observations consist of a multi-frequency, multi-configuration study that includes spectral lines and is sensitive to polarization.  Since the field of view of these observations included N3, we are able to utilize the data to conduct the first in-depth study of N3. In this work we present continuum observations of N3 and the surrounding regions, as well as detailed observations of molecular and hydrogen recombination lines. Detailed discussion of the study as a whole will be presented in future papers. In this paper, we focus on the properties of N3 and examine possible interactions between this source and its surroundings. 

\section{Observations and Data Calibration}
\label{obs}
We carried out multi-frequency, multi-configuration observations of the GC Radio Arc region with the VLA. The WIDAR correlator on the VLA enables the simultaneous observation of a wide spectral bandwidth for sensitive continuum measurements and large number of spectral lines. The observations spanned frequencies from 2 GHz to 49 GHz and used multiple array configurations, thus allowing high resolution observations of the region while still maintaining sensitivity to extended structures (see Table 1). The hybrid arrays of the VLA were used when possible to compensate for the low elevation of the GC from the VLA site. In addition, we utilized data (at 23--25 GHz) from a survey of GC molecular clouds (\citealt{Mills2015}; Butterfield et al. {\it in prep.}). Below we describe the observing and calibration strategy for the ``low frequency'' (2--12 GHz) and ``high frequency'' (30--50 GHz) data separately.

\subsection{Low Frequency Observing and Calibration \\ (2--12 GHz)}

The low frequency observations were conducted using the DnC, CnB, B, and BnA array configurations over a period of several months from May 2013 to February 2014. Multiple pointings were used to cover the brightest portions of the Radio Arc NTFs (see Table 1). The WIDAR correlator was used in 8-bit mode for all observations. All observations were divided into two 1 GHz bands of 8 spectral windows, each consisting of 64 channels. Each set of observations used the same calibrators: 3C286 was used as both the absolute flux and bandpass calibrator, while J1744--3116 served as the phase calibrator. In addition, we observed J1407+2827 as a polarization leakage calibrator as it is a known unpolarized source. The data were then calibrated with the Common Astronomy Software Application (CASA) analysis package \citep{McMullin2007} as provided by the NRAO using the standard processing technique for continuum data. 

\subsection{High Frequency Observing and Calibration \\ (30--49 GHz)}

The high frequency observations were made between May 2013 and January 2014 (see Table 1). The observations were taken using separate 1 GHz sub-bands and consisted of a mixed setup: low resolution spectral windows for continuum observations and high resolution windows for spectral line observations. Continuum spectral windows consisted of 64 channels each, while spectral line observations had varying spectral resolution. Table \ref{spec_image} lists the observed spectral transitions, rest frequencies, velocity resolution, and whether the line was detected in our observations. The standard calibration procedures for high frequency continuum and spectral line observations were followed, including corrections for atmospheric opacity. We utilized 3C286 as an absolute flux calibrator and J1744--3116 as a phase calibrator. J1733--1304 served as the bandpass calibrator. We again observed J1407+2827 as a polarization leakage calibrator, however at these frequencies the calibrator was too weak to detect. Initially using the DnC configuration, we mosaicked a larger field to image the Radio Arc NTFs, but ultimately we were not sensitive to the majority of these structures due to their large angular size. In subsequent observations, we focused only on the field containing N3 to improve our signal to noise on this interesting source. 

\subsubsection{Continuum Imaging}

We created images using the {\it clean} task in CASA: a wide-field image at 5 GHz sensitive to the NTFs and other extended structures in the vicinity of N3 and a set of images focusing on only the compact source N3 at each band of our observations.  To create the wide-field image (Figure 1), we concatenated data from all four array configurations, then created a two-field mosaic using Briggs weighting with a robust parameter of 0.5. In all other images of N3 (see Figure 2), we only used the data from B and BnA arrays with uniform weighting (robust=--2). 

\subsubsection{Spectral Line Imaging}

Table \ref{spec_image} summarizes the spectral lines observed in our spectral line setup. Twelve lines were observed, nine of which were detected in and around N3. Each spectral line was imaged individually and continuum subtracted using the CASA task {\it imcontsub}. All spectral lines, with the exception of \methts{} and \methff{}, were imaged using only the DnC array data at their intrinsic spatial resolution using natural weighting. The images were subsequently smoothed to a spatial resolution of three or four arcseconds in order to improve the image sensitivity. 

The \methts{} and \methff{} transitions at 36 and 44 GHz are well known Class 1 \citep{[collisionally excited;] Morimoto1985,Slysh1994,Sjouwerman2010} maser transitions. Masers are characterized by their point-like spatial distribution and narrow spectral profiles. The \methts{} and \methff{} transitions were imaged with uniform weighting using only the B array data. In addition, self-calibration was performed on a bright maser with minimal neighboring emission in order to increase the signal-to-noise ratio of the observations. The CASA task {\it gaincal} was used to perform phase self-calibration using a model built using {\it clean}. The calibration was applied to the data and the process repeated until the signal-to-noise improvement was no longer significant (2--3 iterations). The phase solutions for this single channel were then applied to all channels in the dataset, and the data imaged with {\it clean} to form the final image.

\section{Results}
\subsection{Properties of N3}

\subsubsection{Continuum structure and size of N3} \label{size}

\begin{figure*}
\includegraphics[scale=0.75]{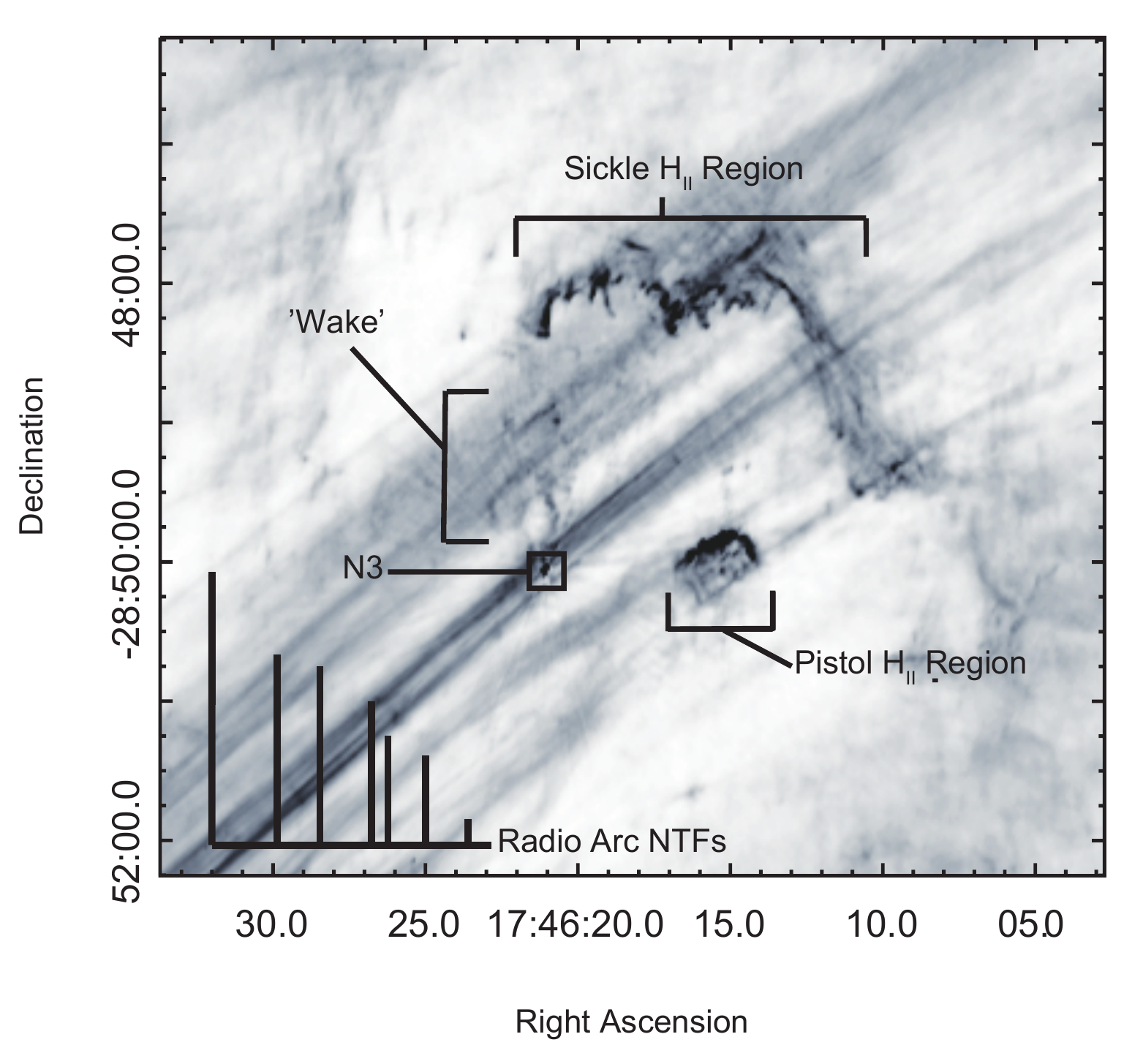}
\caption{5 GHz continuum image of the Radio Arc region, showing the NTFs, Sickle and Pistol \hii{} regions, N3, and the ``wake''.}
\label{finder}
\end{figure*}

\begin{figure*}
\includegraphics[scale=0.9]{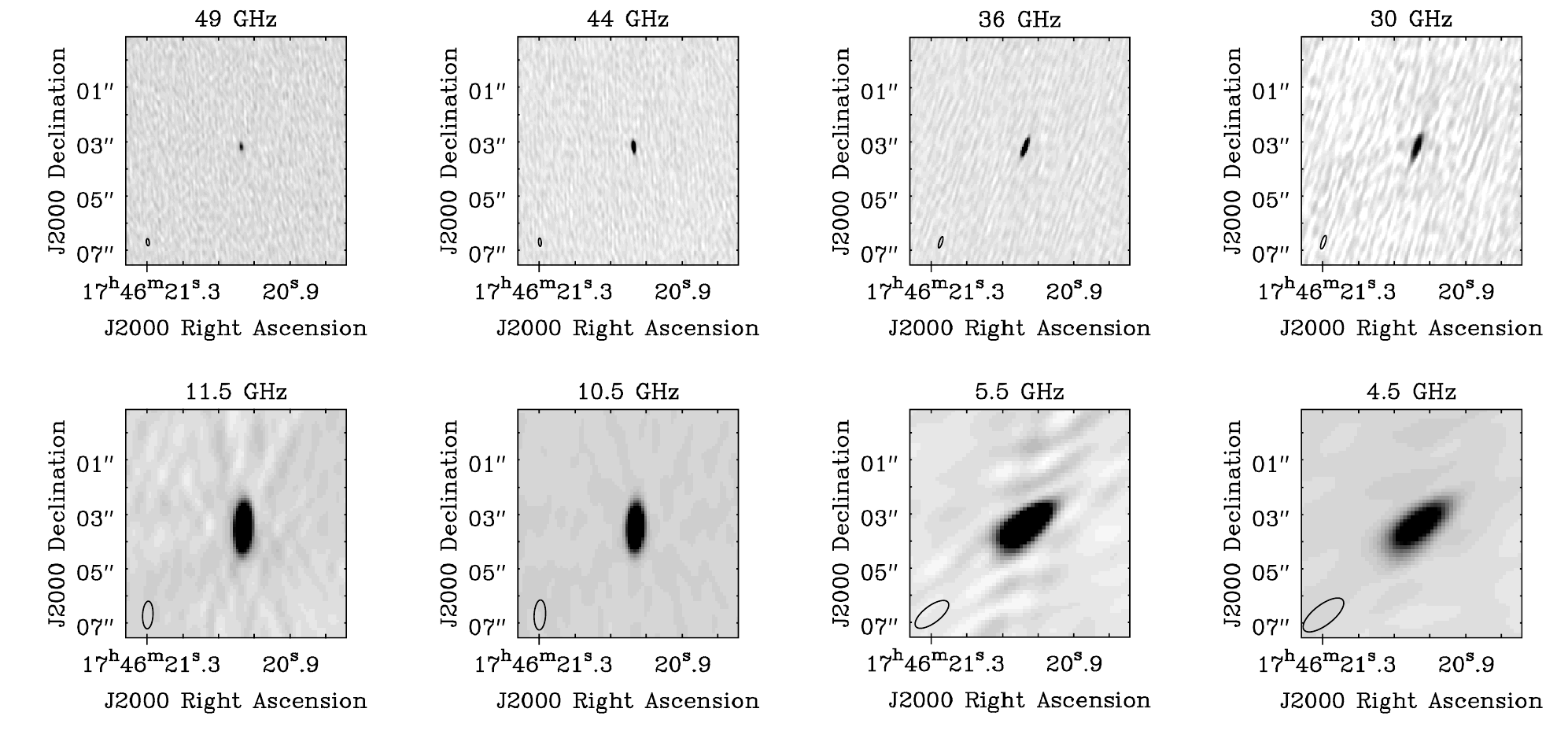}
\caption{Continuum images of N3 from 49 to 4.5 GHz. All images are on the same scale and centered on the position of N3. The synthesized beam for each image is as follows: 49 GHz: 0.26\arcsec{} x 0.11\arcsec{}, 44 GHz: 0.3\arcsec{} x 0.10\arcsec{}, 36 GHz: 0.43\arcsec{} x 0.13\arcsec{}, 30 GHz: 0.52\arcsec{} x 0.15\arcsec{}, 11.5 GHz: 1.01\arcsec{} x 0.38\arcsec{}, 10.5 GHz: 1.10\arcsec{} x 0.43\arcsec{}, 5.5 GHz: 1.48\arcsec{} x 0.62\arcsec{}, 4.5 GHz: 1.80\arcsec{} x 0.74\arcsec{}. The synthesized beam is shown in the bottom left of each image as a black outline.}
\label{continuum}
\end{figure*}

As seen in Figure \ref{finder}, the radio continuum emission from N3 makes it the brightest source observed in the region of the Radio Arc at most frequencies. To examine the structure and size of N3, 
we analyze images made at the following frequencies: 49.0, 44.0, 36.0, 30.0, 11.5, 10.5, 5.5 and 4.5 GHz. Figure \ref{continuum} shows continuum images of N3 created using only data from the B and BnA array configurations with the synthesized beam size in the lower left corner of each panel. The beam sizes range from 0.26\arcsec{}$\times$0.11\arcsec{} (49.0 GHz) to 3.4\arcsec{}$\times$0.80\arcsec{} (4.5 GHz). 

By utilizing only the B and BnA data (see Table \ref{pointings}), we effectively apply a spatial frequency filter that removes the extended emission of the NTFs from the image, which allows us to more easily detect N3. N3 appears to be unresolved at all frequencies (Figure \ref{continuum}). The 49.0 GHz observations (Figure \ref{continuum}, panel 1) provide the highest resolution measurement of this source with a beam size of 260$\times$111 milli-arcseconds (mas). 

To place a stronger constraint on the size of N3, we conducted a Monte Carlo simulation that examined our ability to deconvolve a source from the synthesized beam. In our simulation, we assumed a Gaussian source and SNR of the source, then varied the source size with respect to the synthesized beam. We next attempted to deconvolve the source from the convolution of the source and synthesized beam and determined to what lower limit we could reliably determine a source size. At the SNR of N3 ($\sim25$), we found that a source can be reliably deconvolved from the synthesized beam if the source is $\sim1/4$ the size of the synthesized beam size. However, N3 cannot be deconvolved from the beam, so we place an upper limit of 65$\times$28 mas on the size of N3.

\subsection{Flux Density and Spectrum of N3} \label{spectrum}
The intensity of N3 is brightest in our 10.5 GHz (62 \mjyb) and dimmest at 49 GHz (7 \mjyb); this intensity is nearly an order of magnitude greater than the extended emission of the NTFs and the Sickle and Pistol \hii{} regions in Figure \ref{finder}. Because N3 is bright and detected with a high signal-to-noise ratio in all images, we can study the spectrum of the source across the wide frequency range ($\sim$2$-$36 GHz). Brightness measurements at 44 GHz and 49 GHz were not used due to uncertainties in the flux calibration scale. In order to make an accurate comparison between images, all images were smoothed to a common resolution of 3.5\arcsec{}, which was chosen to match the lowest resolution images (23 GHz). In order to reduce extended structure from the NTFs that might confuse the measurement of N3, we utilized a UV cutoff when making the images.  Forty-eight frequencies were imaged using a 128 MHz bandwidth across our frequency range: 2--4 GHz (5 images, UV cutoff $>$ 40 k$\lambda$), 4--6 GHz (12 images, UV cutoff $>$ 40 k$\lambda$), 10--12 GHz (9 images, UV cutoff $>$ 40 k$\lambda$), 24--26 GHz (12 images, UV cutoff $>$ 20 k$\lambda$), 30 GHz (6 images, UV cutoff $>$ 10 k$\lambda$), and 36 GHz (4 images, UV cutoff $>$ 10 k$\lambda$). For each of these 48 images, we used the maximum intensity of N3 to construct its spectrum. Figure 3 shows the spectrum of N3 across these frequencies. 

\begin{figure*}
\includegraphics[scale=0.75]{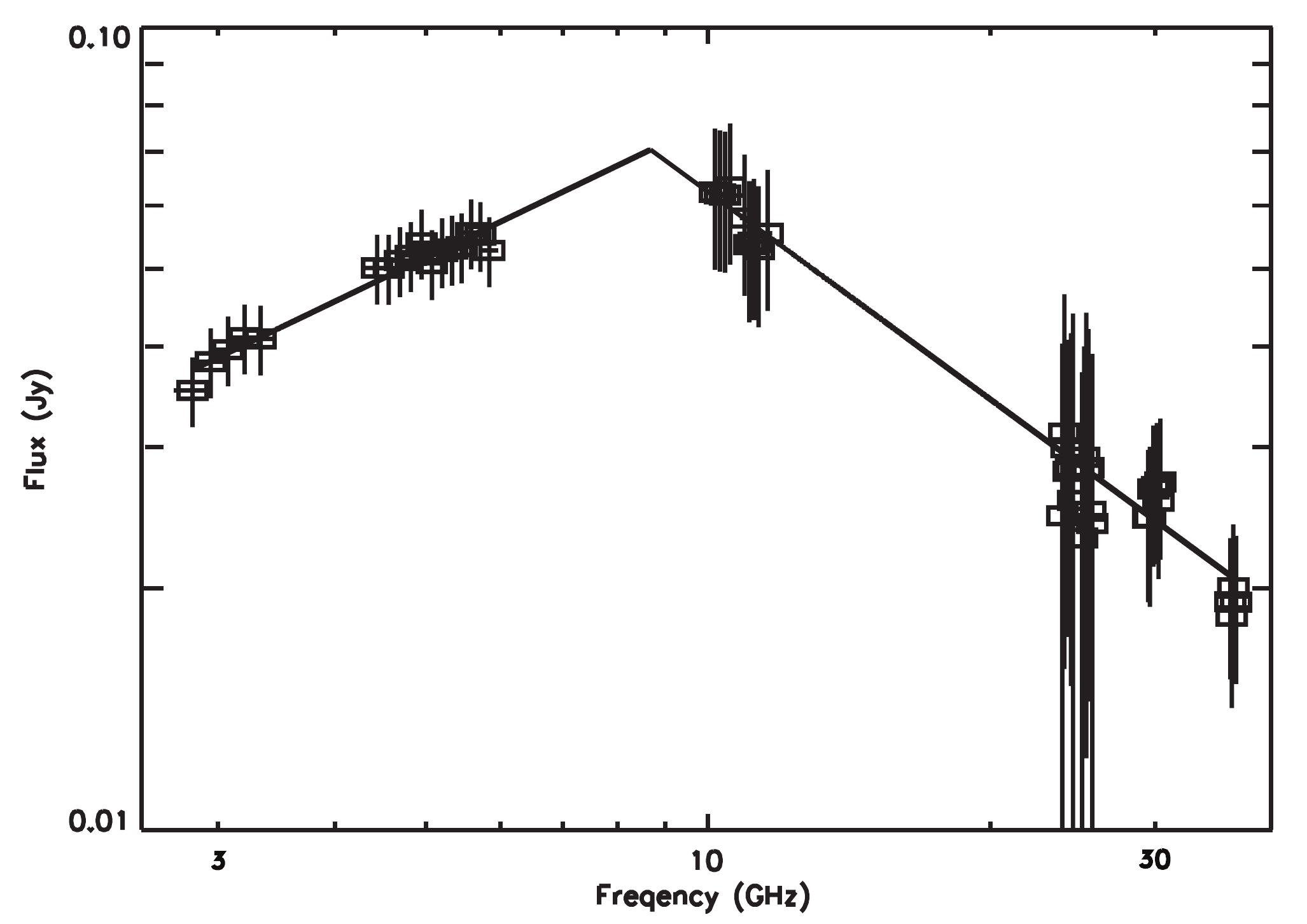}
\caption{Continuum spectrum of N3 from 2 GHz to 49 GHz. The solid line represents a broken power law fit to the data.  At low frequencies (2-6 GHz) the spectral index rises ($\alpha \sim +0.5$) while at high frequencies, (10-36 GHz) the spectrum is falling ($\alpha \sim -0.8$).}
\label{specindex}
\end{figure*}

In order to determine the nature of the spectrum, we compute the spectral index, $\alpha$, defined as $S_\nu \varpropto \nu^\alpha$ where $S_\nu$ is the intensity in mJy and $\nu$ is the observed frequency. To calculate the spectral index, we fit a linear function to the logarithm of intensity versus the logarithm of frequency. Due to the high signal-to-noise ratio of the N3 detection, we assume that errors in the flux of N3 are dominated by the wavelength dependant variation in the background due to the NTFs as opposed to noise in the images. Since each image represents data from a range of frequencies and not a single frequency, we assume a frequency error for each image of the bandwidth of the image (128 MHz). Due to the turnover of the spectrum of N3 at $\sim$8 GHz, we modeled the spectrum of N3 as a broken power law and fitted a spectral index separately above and below the turnover (Figure \ref{specindex}). At low frequencies (2--6 GHz), we assumed a 10\% error in our flux measurements to account for the variable background and assume no variability. Our fit to the low frequency spectrum of N3 results in $\alpha = $+$0.56 \pm 0.13$ with a $\chi$-squared probability of 0.92. 

As will be discussed in Section \ref{var}, N3 appears to be variable over long time scale.  This variability makes determining an accurate spectral index for the high frequency observations difficult since, unlike the low frequency data, our high frequency data were not taken on the same date. In order to produce a good fit the spectral index of N3 at high frequencies, we must assume a 10\% variation in flux in addition to the 10\% error in our flux measurements. Our best fit for the spectral index of N3 is $\alpha \sim $--$0.86 \pm 0.11$ with a $\chi$-squared probability of 0.99. The fits for the high and low frequency spectra indicate that the spectrum turns over at $\sim$ 8.68 GHz.

\subsection{The Environment around N3: the Radio Arc NTFs and "Wake"}

The two most prominent sources near (in projection) to N3 are the Radio Arc NTFs and the ``wake'' structure \citep{Yusef-Zadeh1987} (see Figure \ref{finder}). Our high-sensitivity images reveal greater detail than those of \citet{Yusef-Zadeh1987} and allow us to examine these features and their properties. 
 
\subsubsection{N3 and the Radio Arc NTFs}

\begin{figure*}
\includegraphics[scale=0.5]{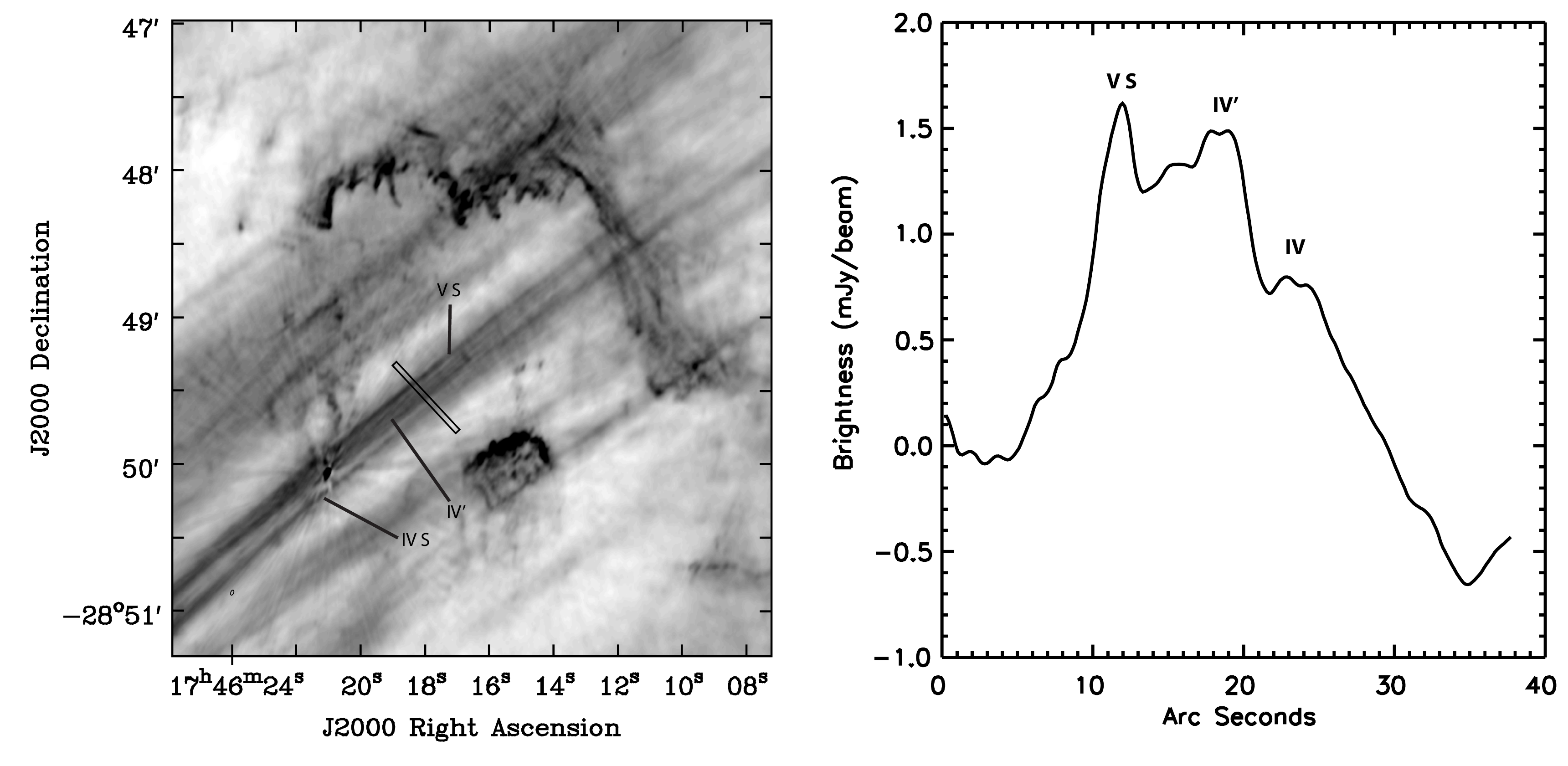}
\caption{Left: 5 GHz image of region surrounding N3. Individual filaments of the NTFs are labeled. Right: Brightness profile of the NTFs as extracted from the black rectangle in the left figure.  The main filaments located near N3 are labeled using the convention of \citet{Yusef-Zadeh1987}.}
\label{ntffind}
\end{figure*}

As first pointed out by \citet{Yusef-Zadeh1987}, N3 appears to be located within the brightest filaments of the Radio Arc NTFs. Though the NTFs are greater than 30 pc in length, they have widths of $\sim 0.1$ pc (Figure \ref{ntffind}, right) \citep[assuming a distance of 8.3 kpc to the GC;][]{Reid2014}. Individual filaments in the NTFs exhibit gentle curvature and appear to intersect at a few locations, such as the intersection of two linear filaments 60\arcsec{} to the south-east of N3. They also display brightness variation along their length, with filaments brightening and fading independent of other nearby filaments. Using the nomenclature of \citet{Yusef-Zadeh1987}, N3 lies along the same line of sight as linear filament IV', which appears to connect with VS in the south and fades to the north of N3 (Figure \ref{ntffind}, left). 

\subsubsection{N3 and the ``Wake''}

\begin{figure*}
\includegraphics[scale=1.0]{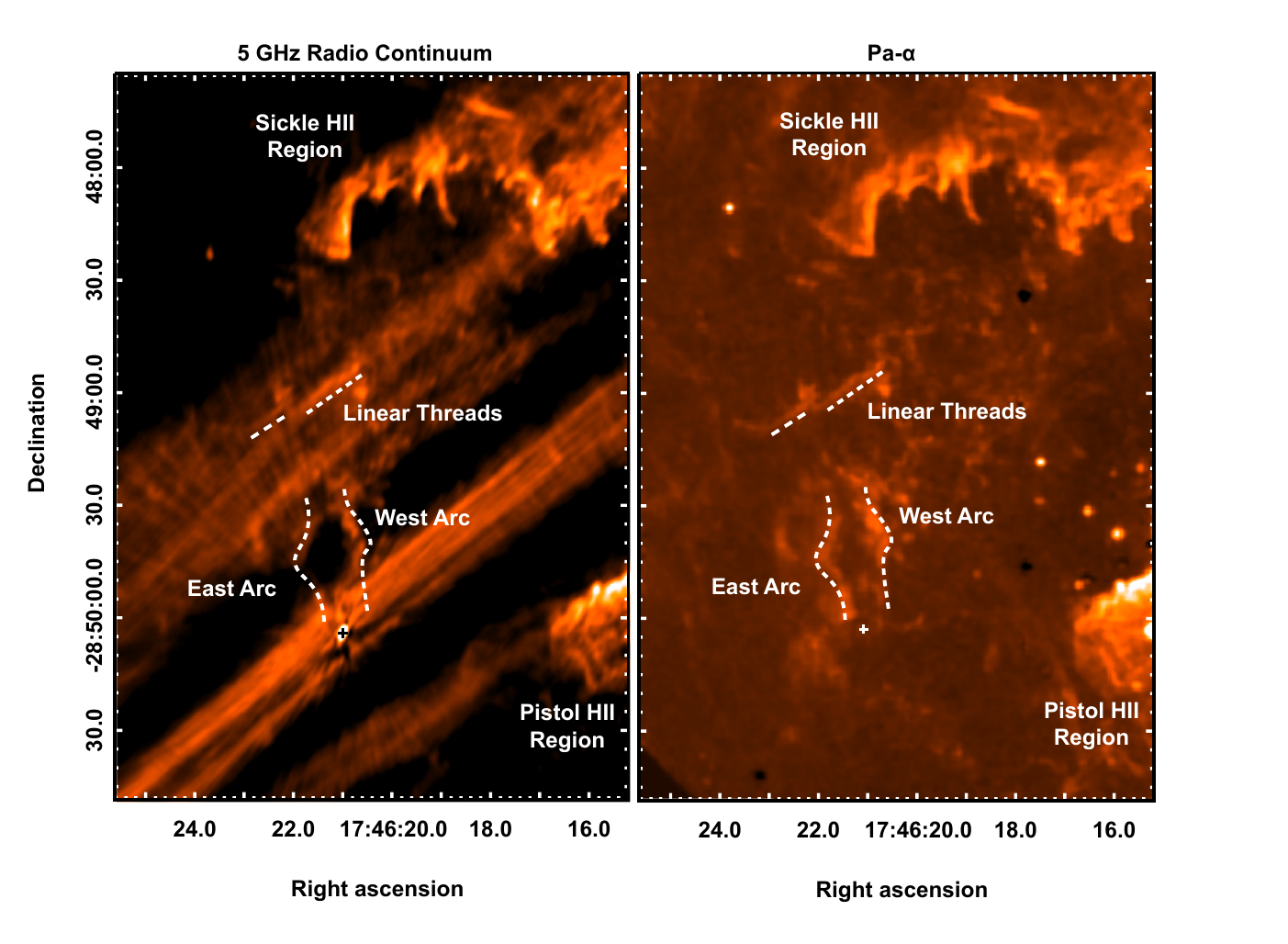}
\caption{Comparison of our 5 GHz continuum observations with HST Pa-$\alpha$ observations \citep{Wang2010}. The ``wake'' is obvious in both the Pa-$\alpha$ observations and the 5 GHz continuum, indicating that the source is thermal in nature. Important features of the ``wake'' are indicated. The position of N3 is marked with the cross. }
\label{palpha}
\end{figure*}

In addition to the NTFs near N3, the other prominent feature in this region is a ``wake'' feature first observed and labeled in \citet{Yusef-Zadeh1987}. The ``wake'' is a series of small, curved filamentary radio continuum features to the north of N3, indicated in Figure \ref{finder}. Compared to the lower sensitivity images of \citet{Yusef-Zadeh1987}, our 5 GHz continuum images reveal a more complicated and detailed structure.  Figure \ref{palpha} is a comparison between our 5 GHz image and the Paschen-$\alpha$ (Pa-$\alpha$) image of \citet{Wang2010} with features of the ``wake'' labeled. These images reveal that the ``wake'' is composed of several separate wisps, filaments, and arcs, all located to the north of N3 (labeled in Figure 5). 

The ``east arc'' and ``west arc'' of the ``wake'' are the two closest Pa-$\alpha$ features to N3. Both arcs are $\sim32$\arcsec{} in length and are roughly oriented symmetrically around a line that would connect N3 to the Western edge of the Sickle HII region. There is no Pa-$\alpha$ emission at the position of N3 itself. To the north of the pair of arcs, a set of thermal linear threads can be seen in both the radio and Pa-$\alpha$ images.  These threads appear to be oriented along the direction of the weak NTFs found in this region. While no bright Pa-$\alpha$ emission is found connecting the linear threads and the pair of arcs, there are numerous faint wisps and clumps found between these two sets of features in the surrounding environment.  

\subsection{The Environment around N3: Molecular Gas}

Surprisingly, strong molecular emission is detected in the region around N3 (see Table \ref{spec_image} for a listing of detected transitions). The molecular emission originates from a compact molecular cloud adjacent (in projection) to N3, hereafter referred to as the ``N3 molecular cloud.''  The two \meth{} maser transitions will be discussed in Section \ref{masers}. We use the observations of the \am{} metastable transitions (\am{}(3,3) through (6,6)) from Butterfield et al. ({\it in prep.}). As shown in Table \ref{spec_image}, no detections of the H56$\alpha$ and H60$\alpha$ lines are made toward this molecular cloud. The lack of radio recombination line emission is also consistent with the lack of thermal radio continuum emission from this extended "N3 molecular cloud" region. 

\subsubsection{Morphology of the N3 Molecular Cloud}\label{molmorph}
 
\begin{figure*}
\includegraphics[scale=1.0]{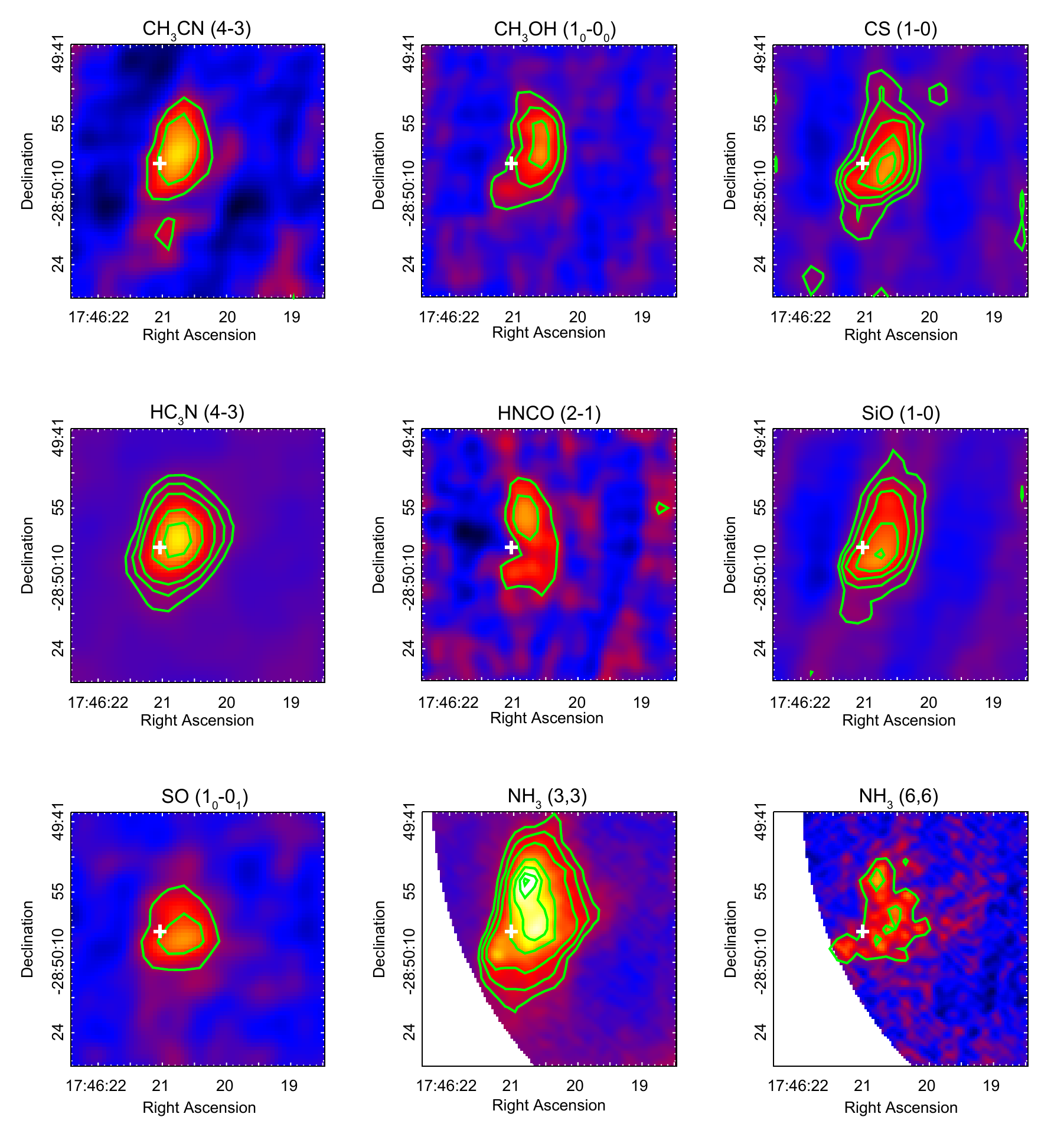}
\caption{Integrated intensity maps of the molecular transitions observed. Contours are at 3, 6, 10, 20, 30, 40, and 50 times the rms noise in each image (\methcy{}: 46 \mjybk{}, \meth{}: 153 \mjybk{}, CS: 188 \mjybk{}, \cyano{}: 98 \mjybk{}, HCNO: 42 \mjybk{}, SiO: 94 \mjybk{}, SO: 100 \mjybk{}, \am{} (3,3): 18 \mjybk{}, \am{} (6,6): 19 \mjybk{}). The position of N3 is marked with a cross.}
\label{molmap}
\end{figure*}

Figure \ref{molmap} presents integrated intensity maps of the observed transitions (the location of N3 is marked with a white cross). The emission from the N3 molecular cloud is roughly the same size in all observed transitions, and is elongated in the north to south direction. In nearly half of the observed transitions, the molecular emission exhibits curvature around the location of N3. Depending on the transition observed, the region of brightest emission is either to the north (HNCO, \am{} (3,3), \am{} (4,4), and \am{} (6,6)) or to the west (\methcy{}, \meth{}, CS, \cyano, SiO, SO, \am{} (5,5)) of N3. The strongest and most widespread emission is from the \am{} (3,3) transition (middle panel, bottom row in Figure \ref{molmap}). 

Figure \ref{cloud} shows the 5 GHz continuum image of the region around N3 overlaid with intensity contours of the \am{} (3,3) transition from integrated -25 to 100 \kms{}. The south-western edge of the \am{} (3,3) emission lies parallel to the NTFs (Figure \ref{cloud}). The eastern edge of the cloud is aligned with the Pa-$\alpha$``west arc.''

\subsubsection{Kinematics of the N3 Molecular Cloud}\label{molkin}

\begin{figure*}
\includegraphics[scale=0.5]{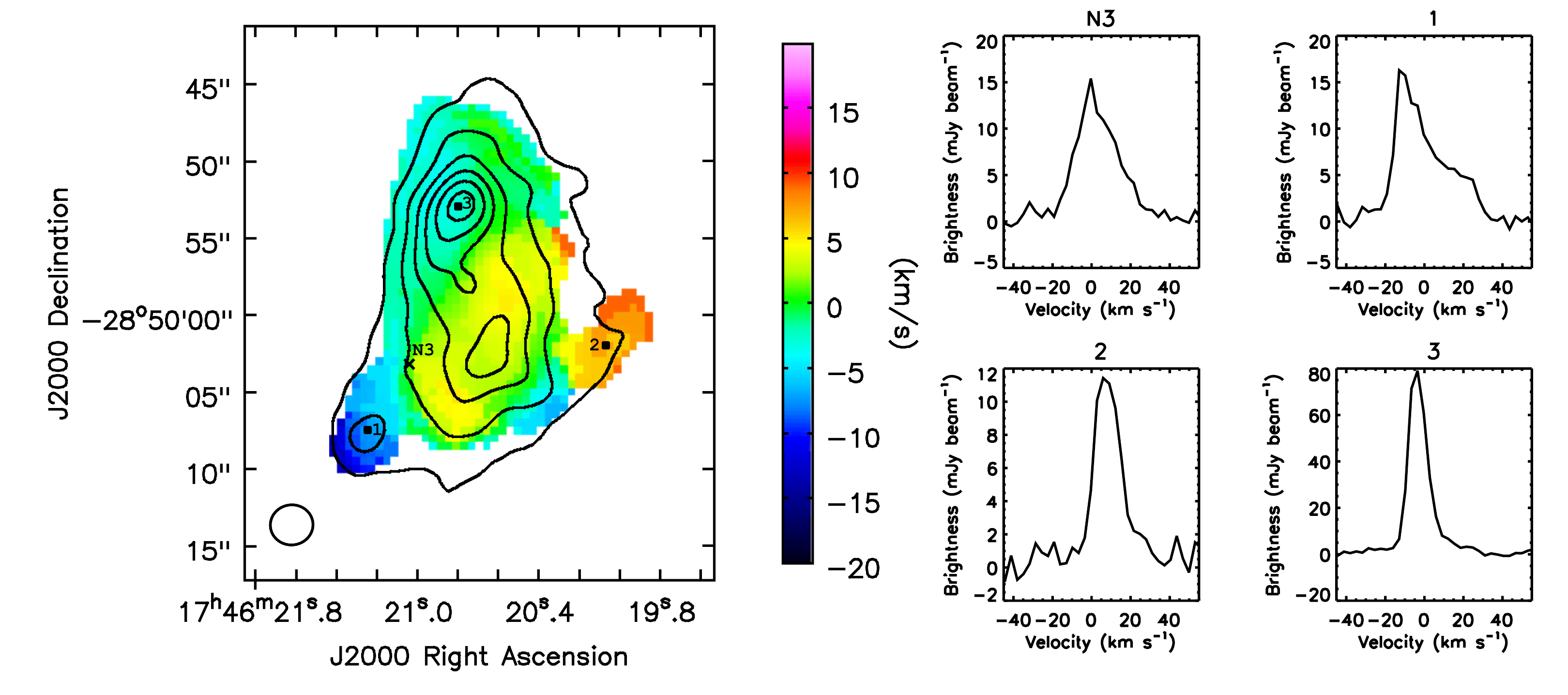}
\caption{Left: Velocity map of \am{} (3,3) emission in the N3 molecular cloud. The contours are 5, 9, 13, 17, 21, and 25 times the RMS surface brightness (40 \mjybk{}) of the integrated \am{} (3,3) image.  N3 is marked by the cross, while three additional positions in the cloud are marked with dots. Right: Spectra of the \am{} (3,3) transition observed in the N3 molecular cloud.  The profiles in this figure correspond to the positions labeled in the left-hand figure.}
\label{kinmaps}
\end{figure*}

\begin{figure*}
\includegraphics[scale=0.5]{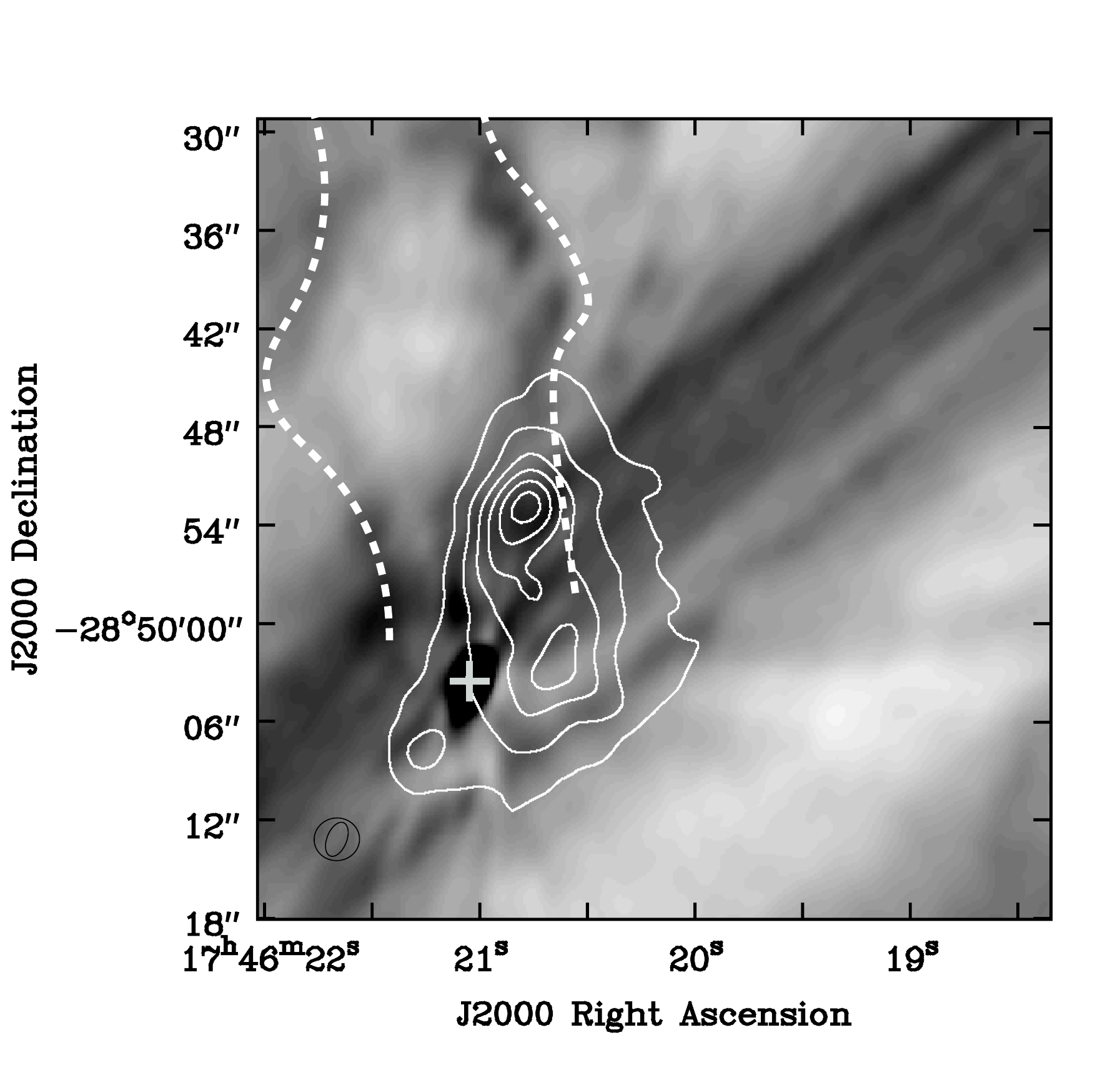}
\caption{5 GHz continuum image (greyscale) overlaid white contours showing \am{} (3,3) emission integrated between velocities of -25 and 100 \kms{}. Contour values are 5, 9, 13, 17, 21, and 25 times the RMS noise value of 40 \mjybk{}. The dashed white lines are the outlines of the "arcs" shown in Figure \ref{palpha}. The cross denotes the position of N3.}
\label{cloud}
\end{figure*}

Figure \ref{kinmaps} shows the distribution of peak velocity derived from a single Gaussian profile of the \am{} (3,3) emission within the N3 molecular cloud, as well as \am{} (3,3) spectra towards N3 and three selected positions in the cloud. A velocity gradient exists in the cloud, with the velocity roughly increasing from east to west across the cloud. This gradient is strongest between positions 1 and 2 with a velocity gradient of $\sim 30$ \kms{} pc$^{-1}$.  The gradient between position 1 and position 3 is $\sim 15$ \kms{} pc$^{-1}$ while the gradient between position 2 and position 3 is $\sim 21$ \kms{} pc$^{-1}$. These velocity gradients are significantly larger ($\sim$2 times) than the velocity gradients measured in several other Galactic center clouds \citep[i.e.,][]{Lang1997,Lang2001,Mills2015}.

The velocity profiles shown in Figure \ref{kinmaps} show that the molecular lines in the N3 molecular cloud have broad line widths, particularly in the south-east region of the cloud. Near position 1 and N3, emission is detected between $\sim$ -20 to +20 \kms{}.  The profiles also show unusual structure; this wide structure may represent the superposition of multiple velocity components (multiple profiles in this region may also explain the unusually large velocity gradients, derived from a single broad profile at one of the locations). The lines in the western and northern portions of the cloud show narrower, single-peaked, Gaussian-like profiles with FWHM of $\sim$ 10 to 15 \kms{}. We also note that no absorption is seen in the spectral profile toward N3.

\subsubsection{Ammonia Temperature}\label{ammtemp}

By observing multiple metastable (J=K) transitions of \am{}, we are able to measure the temperatures of the N3 molecular cloud. \am{}, due to its low dipole moment, is able to reach thermal equilibrium with H$_2$ relatively quickly, while the short lifetimes of the non-metastable (J$\neq$K) states leave most molecules in the metastable states. By comparing the column densities of the metastable states, a rotational temperature can be determined that provides a good approximation to the kinetic temperature of the cloud \citep{Morris1973,Huettemeister1993}. 

When calculating the temperature using the metastable \am{} transitions, it is important to know whether the observed transitions are optically thick or thin. For an optically thick cloud, the hyperfine satellite lines of the transition will be enhanced relatively more than the main hyperfine line. While strong hyperfine features were observed in the \am{} (1,1) and (2,2) transitions, no hyperfine structure is observed in any of the higher metastable transitions, thus we conclude that the N3 molecular cloud is optically thin for the \am{} (3,3), (4,4), (5,5), and (6,6) transitions.   

To determine the temperature of a cloud, two transitions of either para (K$\neq$3n) or ortho (K=3n) \am{} are needed. For optically thin clouds, the column density can be calculated from the velocity integrated brightness temperature by:

        \begin{equation} N (J,K) = \frac{1.55\times10^{14}\, \mathrm{cm}^{-2}}{\nu}\frac{J(J+1)}{K^2}\int T_{mb}\, \mathrm{dv}\,,
        \label{eqa}
        \end{equation}

as in \citet{Huettemeister1993} and \citet{Mauersberger2003}. We utilized the \am{} (3,3), (4,4), (5,5), and (6,6) transitions to construct maps of column density from the integrated intensity images of each transition. These maps were then used in pairs to determine the average rotational temperature $T_{rot}(J,J')$ of the cloud using the Boltzmann equation: 

	\begin{equation} \frac{N (J,K=J)}{N(J',K'=J')} = \frac{g_{op}(K)}{g_{op}(K')} \frac{(2J+1)}{(2J'+1)} exp \left( \frac{kT_{rot}(JJ')}{-E} \right)
	\label{eqc}
	\end{equation}

where $E$ is the energy relative to the ground state and $g_{op}(K')$ is the statistical weight for para ($g_{op}(K=3n)=1$) or ortho ($g_{op}(K\neq3n)=2$) transitions.  To determine rotational temperatures within the N3 molecular cloud we examined both the \am{} (3,3) and (6,6) ortho transitions and the \am{} (4,4) and (5,5) para transitions. The average \am{} (3,3) -- (6,6) temperature ($T_{rot}(3,6)$) over the N3 cloud is 81$^{+8}_{-7}$ K. The \am(4,4) -- (5,5) ($T_{rot}(4,5)$) temperature is much higher than T$_{36}$, with an average of 170$^{+140}_{-50}$ K averaged over the N3 molecular cloud. 

\subsubsection{36 and 44 GHz Methanol Maser Lines}\label{masers}

\begin{figure*}
\includegraphics[scale=1.0]{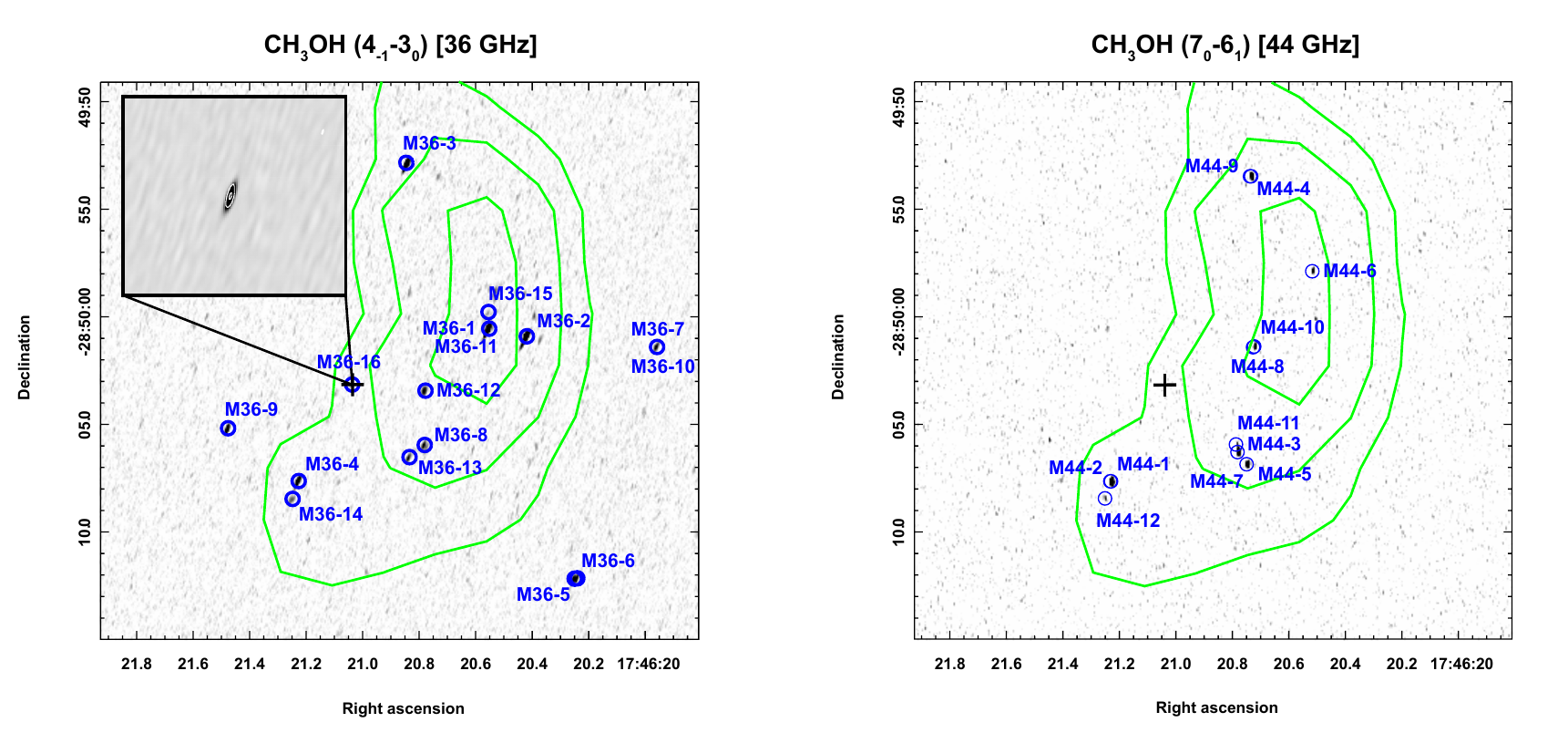}
\caption{Finding charts for detected 36 GHz (Left) and 44 GHz (Right) \meth{} masers in the N3 cloud. Greyscale shows maximum intensity maps of each \meth{} transition while the green contours show the thermal \meth{} (1$_0$-0$_0$) transition at 3, 6, and 10 times the rms (153 \mjybk{}) of the integrated intensity map.  N3 is marked by the black cross. The inset shows the 36 GHz continuum image of N3 (grey scale) with 11 \kms{} \methts{} contours overlain at 4 and 8 time the channel RMS (RMS = 4.5 mJy/beam, channel width = 1 \kms{}).}
\label{maserfinder}
\end{figure*}

We detect over a dozen compact, point-like sources in the 36 GHz \methtst{} and 44 GHz \methfft{} transitions of \meth{} (Figure \ref{maserfinder}). These transitions are well known to give rise to collisionally excited, Class I masers and to trace shocks \citep{Morimoto1985,Slysh1994,Sjouwerman2010}. To examine the possibility that the point sources in the N3 clouds are maser sources, we examined the properties of each source.

In order to have uniform selection criteria for characterizing the point sources, we have utilized the source detection algorithm {\it clumpfind} \citep{Williams1994}. {\it Clumpfind} distinguishes sources that partially overlap in position or velocity. The algorithm identifies local maxima, then examines the emission surrounding the maxima both spatially and spectrally to determine the boundaries of the source. No assumptions about the clump geometry, either spatially or spectrally, are made during processing by the algorithm. {\it Clumpfind} produces a list of sources with uniform selection criteria. The output of {\it clumpfind} was used to construct a catalog of sources for both the 36 and 44 GHz transitions. The {\it clumpfind} analysis yields a total of sixteen 36 GHz \methts{} and twelve 44 GHz \methff{} point sources. Since the sources are not centered on the field, a primary beam correction was applied to the intensities of the detected sources.  All maser candidates possess a brightness greater than six times the rms noise in their spectral channel. The point sources detected at 36 GHz and 44 GHz were then compared to determine any coincidences between the two transitions. Three point sources displayed both lines.

\begin{figure*}
\includegraphics[scale=0.8]{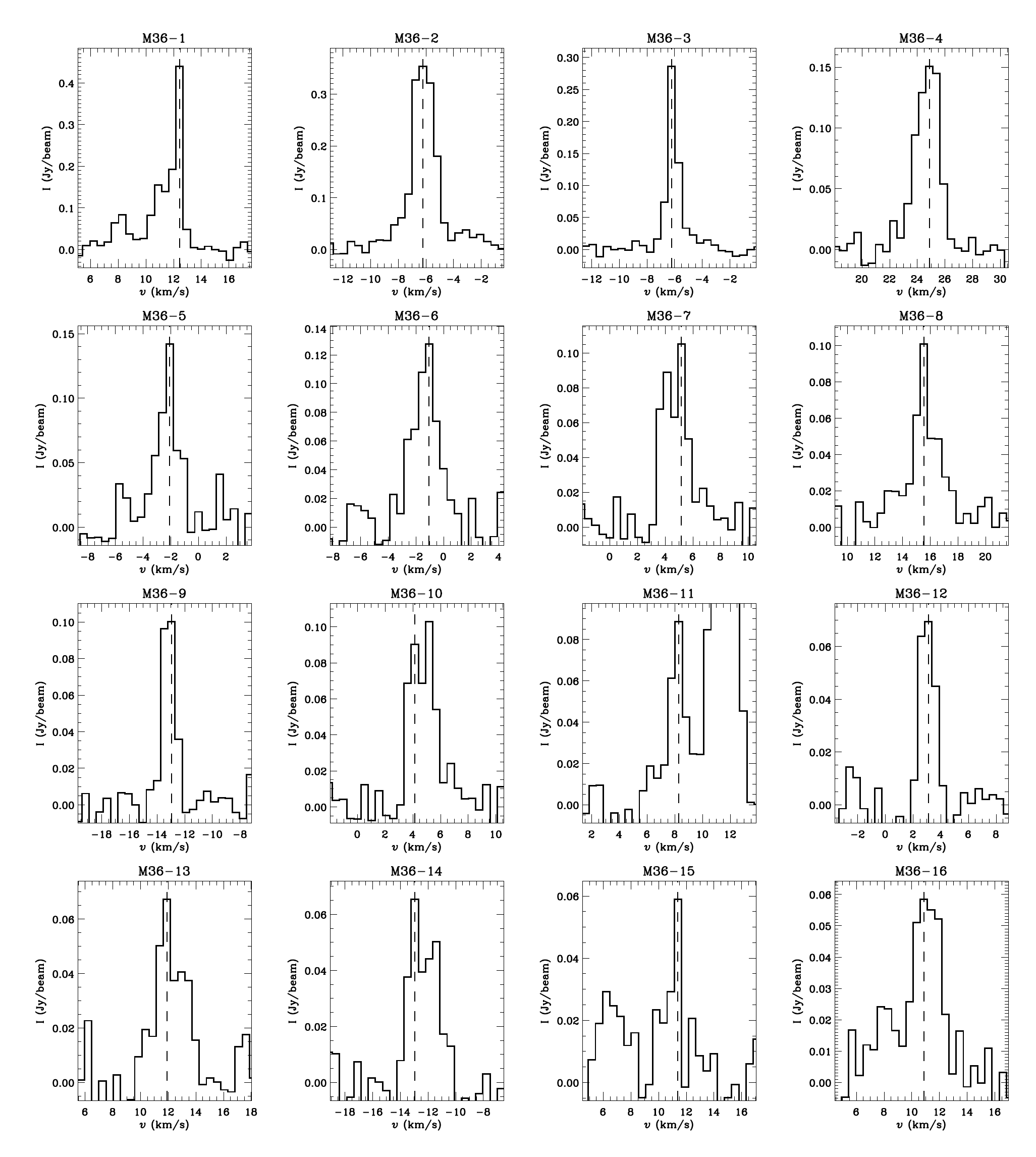}
\caption{Spectra for each detected 36 GHz \methts{} Maser. The dashed line marks the velocity for the detected maser.}
\label{36ghz_spec}
\end{figure*}

\begin{figure*}
\includegraphics[scale=0.8]{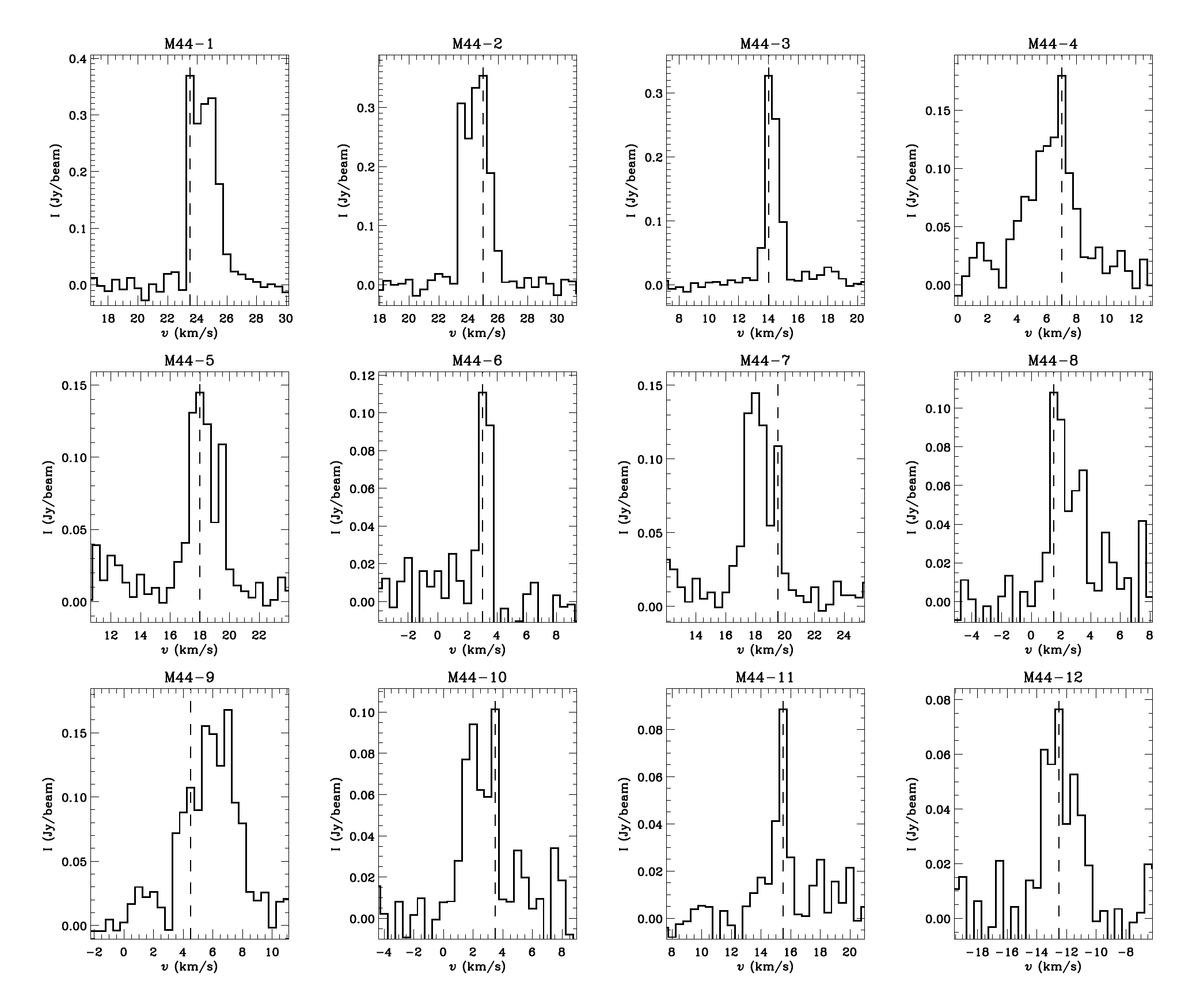}
\caption{Spectra for each detected 44 GHz \methff{} Maser. The dashed line marks the velocity for the detected maser}
\label{44ghz_spec}
\end{figure*}

The properties of all point sources can be seen in Tables \ref{36Masers} and \ref{44Masers} which correspond to the \methts{} and \methff{} transitions, respectively.  These tables provide: (1) Catalog Number, (2) Galactic Name, (3) Right Ascension (HH:MM:SS.s), (4) Declination (DD:MM:SS.s), (5) velocity (\kms), (6) FWHM (\kms), (7) Peak Brightness (Jy beam$^{-1}$), (8) Flux (Jy), (9) Brightness Temperature (K), and (10) any counterpart in the other transition. In addition to the source catalog, spectra are also presented for each source in Figures \ref{36ghz_spec} and \ref{44ghz_spec}. The spectra do not have velocity cutoffs for the edges of the line, thus faint sources may not be the strongest peak in their spectra if a brighter source was located nearby.  To aid in identification of weak sources near brighter sources, the central velocity of each source is indicated by the dashed line in each spectrum. Additionally, a finding chart of the detected sources can be found in Figure \ref{maserfinder}. The brightness temperatures of the detected sources are all in excess of 900 K. Since this temperature is nearly half an order of magnitude greater than the \am{} temperature of the cloud, we conclude that these sources are not thermally excited and must be masers. 

The majority of 36 GHz masers and all of the 44 GHz masers lie along an arc with a center offset by $\sim$ 2\arcsec{} to the north from the position of N3. This arc of maser emission also contains the three brightest regions of \am{} (3,3) emission, though the masers are not clustered around the \am{} (3,3) clumps. This arc is positionally coincident with the brightest regions of the other molecular transitions, though the intensity of the masers does not correlate with the intensity of the other observed transitions. The \meth{} masers have velocities ranging from --13 \kms{} to +25 \kms{}, consistent with the full range of velocities observed in the other spectral lines.     

While there are many masers within the cloud, the weakest detected 36 GHz \meth{} maser in the cloud is perhaps the most interesting. M36-16 is positionally coincident with continuum emission from N3 to within the positional accuracy of our observations (Figure \ref{maserfinder} inset). Apart from its interesting location, maser M36-16 resembles the other masers within the N3 molecular cloud in its velocity and velocity width. No 44 GHz counterpart to M36-16 is detected. 

\section{Discussion}       
\label{dis}

\subsection{Nature of Continuum Emission from N3}

\subsubsection{Size}

As discussed in Section \ref{size} and Figure \ref{continuum}, N3 is unresolved at all observed frequencies. Our highest resolution data at 49 GHz has a resolution of 260$\times$110 mas for the angular size of N3. This angular size corresponds to 0.011$\times$0.005 pc (2200$\times$1000 AU) if N3 is located at the distance of the GC \citep[8.3 kpc;][]{Reid2014}. However, our Monte Carlo simulation (Section \ref{size}) shows that the upper limit to the size of N3 is $\sim 1/4$ of the synthesized beam size at our SNR of $\sim25$, or $65\times 28$ mas, corresponding to a linear size of 0.0028$\times$0.0013 pc (550$\times$250 AU).  The size of N3 itself constrains the nature of the sources by limiting physical counterparts to very compact sources as discussed in Section \ref{nature}.

\subsubsection{Spectrum}

To understand the unusual source N3 we must understand the emission mechanisms at work within the source. As discussed in Section \ref{spectrum}, the high frequency (10$-$49 GHz) spectrum of N3 falls with $\alpha = -0.86 \pm 0.11$ while the low frequency (2$-$6 GHz) spectrum rises with $\alpha = +0.56 \pm 0.13$. 

The high-frequency spectrum of N3 cannot be produced by thermal, free-free emission, which typically has a spectrum of $\alpha \sim -0.1$.  Assuming a power law distribution of relativistic electrons, an optically thin synchrotron spectrum will have a spectral index $\alpha = 0.5 (1-\delta)$ where $\delta$ is the power law index for the electron distribution. Consequently, at high frequencies, an optically thin synchrotron spectrum is the best interpretation of the observed spectrum of N3.

Two possibilities exist to explain the spectrum at low frequencies: 1) free-free absorption of the synchrotron source and 2) synchrotron self-absorption \citep{deBruyn1976,Artyukh2001}. Free-free absorption within a synchrotron source would act to flatten the spectral index of the synchrotron source over a narrow range of frequencies. However, the low frequency spectrum of N3 maintains a power law over a full dex in frequency. A free-free absorption from a uniform foreground source would not produce a power law spectrum over this large of a wavelength range. Additionally, a mixed synchrotron and thermal gas with internal free-free absorption cannot reproduce both the high and low frequency spectrum of N3. A non-uniform free-free absorbing medium could be designed to fit the observed spectrum, however the free-free absorbing material must be very compact ($<30$ mas) in order for the thermal radio continuum emission to not be observed. Therefore, we conclude that free-free absorption of a optically thin synchrotron source not likely to produce the observed spectrum of N3.

Self-absorbed synchrotron emission from a uniform source has a theoretical spectral index of +2.5, much steeper than the low frequency spectrum of N3. However, a self-absorbed synchrotron spectrum can be flattened if the source is non-uniform, with the magnetic field decreasing from the center to edge of the emission region \citep{deBruyn1976,Artyukh2001}.  

\subsubsection{Variability} \label{var}

Although no significant short-term variability is observed from N3 within the timespan of our observations, N3 appears to be variable on long time scales. \citet{Yusef-Zadeh1987} reported a peak intensity of 13.6 mJy beam$^{-1}$ at 4.8 GHz with a bandwidth of 12.5 MHz and a beam of 2.58"$\times$1.93" in observations taken between 1982 and 1985. To directly compare our data with the results of \citet{Yusef-Zadeh1987}, we created an image from our observations using the same center frequency, bandwidth, and synthesized beam. In our image, we find an intensity of 45.5 mJy beam$^{-1}$, over a factor of three greater than the intensity measured by \citet{Yusef-Zadeh1987}. Contamination by the NTFs would affect both observations equally and thus cannot account for the difference. 

\citet{Lang1997} observed N3 at 8.3 GHz using data taken by the VLA during 1992 and 1993. In their image, the peak intensity of N3 is 15.4 \mjyb{}. While we do not have 8.3 GHz data to directly compare with the \citet{Lang1997} observations, if we assume that the spectral index of N3 has remained the same we can compare the intensity at 8.3 GHz to the intensities observed in this work and in \citet{Yusef-Zadeh1987}. Using the spectral index of +0.56 determined in Section \ref{spectrum}, we estimate that the 4.8 GHz intensity of N3 in 1993 was $\sim 11$ \mjyb{}. While the inferred 4.8 GHz intensity of N3 in 1993 is comparable to the measured 4.8 GHz intensity in 1982-85, N3 has undergone significant brightening between 1993 and 2013. We conclude that N3 is likely variable over decade-long time scales, however improved observations of radio variability are clearly needed.

\subsection{Relation between N3, the N3 molecular cloud, NTFs, and the ``Wake''}

\subsubsection{Is the N3 molecular cloud located at the Galactic Center?}\label{cloudGC}

In order to better constrain the location of the N3 molecular cloud, we can look for similarities in the physical properties between the N3 molecular cloud and molecular clouds known to reside in the GC. While clouds within the Galactic disk have narrow line widths (2-10 \kms) and low temperatures, clouds in the Central Molecular Zone (CMZ) typically have larger linewidths \citep[15--50 km s$^{-1}$;][]{Bally1987} and high temperatures \citep[50--300 K;][]{Huettemeister1993}. Figure \ref{kinmaps} shows that the molecular gas observed in the N3 molecular cloud has linewidths of 10--30 \kms{}, consistent with most molecular clouds in the CMZ. The rotational temperatures of the N3 molecular cloud, as determined in section \ref{ammtemp}, are $T_{rot}(3,6)$=81 K in and $T_{rot}(4,5)$=165 K. These rotational temperatures are comparable to the temperatures typically observed in CMZ clouds \citep{Huettemeister1993} and are atypical of molecular clouds in the Galactic plane. Due to the broad line widths and high temperatures of the cloud, we conclude that the N3 molecular cloud likely lies within the CMZ. A location in the GC implies a physical size of the cloud of 0.8 pc $\times$ 1.04 pc, assuming a distance of 8.3 kpc.  

\subsubsection{Origin and Nature of the N3 Molecular cloud?}\label{originN3cloud}

Assuming (as discussed above) that the N3 molecular cloud is located at the GC, then it has a relatively compact physical size ($\sim$1 pc). Therefore, it is useful to explore its origin and possible connection to surrounding molecular material. As described earlier, both N3 and the N3 molecular cloud lie in projection near to the ``25 km/s'' molecular cloud originally studied by \citet{Serabyn1991}.  In their Figure 2, emission is evident in the upper right panel (velocities of --5 to 10 \kms{}) in the CS (J=3-2) emission line that is coincident with the position of N3 and relatively compact in nature. Faint emission at velocities near 25 \kms{} is also observed from this region in these early single-dish studies. It is therefore plausible that this compact cloud is physically related to a much larger molecular cloud structure in this region. The N3 cloud lies $\sim$4 pc in projection away from the massive Quintuplet stellar cluster, which is capable of ionizing the adjacent and extended Sickle HII region \citep{Lang1997,Figer1999} at a large distance. Yet, there is no radio continuum arising from the edge of the N3 molecular cloud in our data.  However, it is possible to explain the lack of ionization in the N3 molecular cloud if the ionizing photons from the Quintuplet were blocked by intervening gas. Several clumps of gas have been observed between the N3 molecular cloud and the Quintuplet (see \citet{Serabyn1991} and Butterfield et al. 2016, in preparation) which may be shielding the N3 molecular cloud from ionizing radiation. 

\subsubsection{Physical arrangement of N3, the N3 molecular cloud, and the NTFs}\label{intcloud}

As Figure \ref{molmap} shows, there is molecular emission present in all spectral lines at or near the location of N3. The N3 cloud clearly encompasses the position of N3. If N3 were located on the far side of the N3 molecular cloud, we would expect to see absorption in the spectral line profiles of molecular gas. However, as is shown in the spectral line profile in Figure \ref{kinmaps}, no absorption is seen at the position of N3. Since we do not see absorption and we do see line emission toward N3 from most observed transitions, it is likely that N3 lies on the near side of the N3 molecular cloud. 

As described earlier, maser source M36-16 is located at the same location as N3. We calculate the probability of a chance superposition of N3 with any of the masers within the cloud to be 0.3\%. If the \methts{} transition were inverted in the foreground gas of N3, the continuum emission from N3 would be amplified at the local cloud velocity. The low probability of a chance superposition indicates that N3 is at least partially embedded in the cloud. If N3 is located just within the surface of the cloud, we could account for weak amplification of the \methts{} transition, while the background emission would dominate any absorption in the other molecular transitions.   

The morphology of the N3 molecular cloud is suggestive of a possible association between N3 and the N3 molecular cloud. The molecular line emission from the cloud appears to exhibit a curved morphology around the location of N3 (Figure \ref{molmap}). This curvature may indicate that N3 is interacting physically with the molecular cloud. Going further, in order to explore the physical arrangement between the NTFs and the N3 molecular cloud, one can examine the morphology of the observed spectral line emission.  The southern edge of the N3 molecular cloud is aligned with the NTFs, while the western edge is roughly perpendicular to the NTFs. In the north, the molecular emission extends past the NTFs where the ``west arc'' of the ``wake'' intersects the NTFs. (Figure \ref{cloud}). While the observed alignment of the southern edge of the molecular emission and the NTFs may be coincidental, the possibility exists that the N3 molecular cloud is bounded by the NTF at this location. \citet{Serabyn1994} noted that molecular gas in the ``25 \kms{}'' molecular cloud (to the north of the N3 molecular cloud) appears to be elongated along the NTFs, which suggests a possible interaction. The morphology of the N3 molecular cloud may indicate a similar interaction between the N3 molecular cloud and the NTFs.

\subsubsection{Relationship between N3 and the ``Wake''}\label{intwake}

N3 is located to the south of the ``east arc'' and ``west arc'' of the wake.  The ``east arc'' and ``west arc'' both appear to extend from near the position of N3 northwards away from the N3 molecular cloud (Figure \ref{palpha}). When considering both features, the curvature of both the arcs 22\arcsec{} (0.9 pc) north of N3 is reminiscent of a shell with a diameter of 12\arcsec{} (0.5 pc).  The morphology of the arcs and shell like feature may be evidence of a past energetic event. 

The ``linear threads'' of the ``wake'' lie between the ``arcs'' and the Sickle \hii{} region to the north and are aligned with a few of the NTFs that are located in this region. We conclude that they are likely associated with the NTFs, given their alignment along the filaments. We do not see evidence that the ``linear threads'' are associated with N3. 

\subsection{What is the nature of N3?}\label{nature}

With the wealth of new knowledge concerning N3, we revisit existing hypotheses for its physical nature. These include: 

(1) UC\hii{} Region: An Ultra Compact \hii{} region typically possesses a flat radio continuum spectrum and would be detectable in the observed radio recombination line images and the Pa-$\alpha$ emission of \citet{Wang2010}. The continuum spectrum of N3 is unambiguously non-thermal in nature, and no emission is detected in radio recombination or Pa-$\alpha$ lines. This allows us to rule out an UC\hii{} region as the physical counterpart to N3. 

(2) Young supernova: The radio emission expected from a young supernova (SN) could account for the spectrum of N3. However, a young SN would be expected to expand rapidly. N3 was first observed in 1982, while our observations were completed in 2014.  If we assume a SN occurred in the GC shortly before the observations of \citet{Yusef-Zadeh1987}, the expansion velocity would be less than 38 km/s (assuming 250 AU) if the SN were to remain unresolved in our high resolution observations. Additionally, the shock generated by a SN within the N3 molecular cloud would likely generate X-ray emission. In Chandra observations of the region (discussed in detail below) no X-ray emission is seen. Due to the low upper limit on the expansion velocity and the lack of X-rays, can rule out a young SN within our galaxy.

(3) Foreground Active Star: A foreground active star, such as an RS CVn star could account for the non-thermal radio emission from the continuum source. However, the emission from active stars is typically circularly polarized \citep{Mutel1987} whereas we detect no polarization from N3.  Also, such an object would have to be relatively nearby ($<100$ pc) to account for the observed radio flux. An RS CVn star at that distance should be easily detectable in the optical, however, no optical counterpart is observed in the Second Digitized Sky Survey \citep{McLean2000} nor in the 1.9 micron observations of \citet{Wang2010}.

(4) Active Galactic Nucleus: The bright self-absorbed synchrotron emission and very compact size are characteristic of AGN. While the physical properties of N3 itself seem well matched to the AGN hypothesis, the lack of absorption in our observations suggests that N3 lies in front of most of the molecular gas, and thus within the Galaxy. Also, if N3 were an AGN, the apparent morphological association of the molecular cloud with N3 would then have to be an accidental coincidence.      

(5) Micro-quasar: Micro-quasars are a type of X-ray binary in which material from a companion star is accreted by a stellar mass black hole. Micro-quasars can produce many of the observable features of N3, including self-absorbed synchrotron emission and point-like emission \citep{Rodriguez1995,Dhawan2000,Soria2010}. A micro-quasar origin for N3 would also fit well with the observed radio flux. However, if N3 were a micro-quasar, the bright radio emission would indicate that bright X-ray emission should also be present \citep{Remillard2006}. Chandra images of the region (Obs ID 14897) show no source at the position of N3. The column density along the line of sight to N3, as determined using the WebPIMMS utility \footnotemark[2]\footnotetext[2]{https://heasarc.gsfc.nasa.gov/cgi-bin/Tools/w3pimms \\ /w3pimms.pl}, is $N_H\sim$ $1.2\times10^{22} $ cm$^{-2}$. Using the non-detection in the Chandra observation and this value of $N_H$, we place an upper limit on the unabsorbed X-ray flux from N3 of $\sim10^{-15}$ erg cm$^{-2}$ s$^{-1}$. The observations with the VLA and Chandra are separated by 28 days. The observation of strong radio emission with no detectable X-ray counterpart is unlike any observed micro-quasar, thus we conclude that the micro-quasar hypothesis is also unlikely.

(6) Other: Since all of the above hypothesis for N3 are essentially ruled out, we must consider that N3 represents a more exotic object. Could N3 possibly represent a black hole larger than the stellar mass black holes of micro-quasars? To explore this possibility, we attempted to constrain the possible mass of N3 using the fundamental plane of black hole activity \citep[Equation 5,][]{Merloni2003}. Unfortunately, the errors in this relation are too large to provide a meaningful constraint on the nature of N3. However, we also examined Table 1 of \citet{Merloni2003}, which lists the X-ray and 5 GHz radio luminosities of Galactic black holes and AGN used to derive the fundamental plane of black hole activity. In this table, every black hole source possessed an X-ray luminosity greater than its radio luminosity. This situation is reversed for N3, which has a radio luminosity greater than the upper limit of its X-ray luminosity. Because of the low X-ray luminosity, we conclude that N3 is unlikely to be a typical hard state black hole.

The sample of black holes examined by \citet{Merloni2003} excludes blazars. In blazars, the accretion powered radio jet is directed nearly along the line of sight to the observer. Relativistic beaming is then able to increase the intensity of the observed jet with respect to the X-ray emission of the hot corona. While the increase in brightness is heavily dependent upon the orientation of the beam to the observer and the speed of the relativistic particles, an amplification of more than two orders of magnitude is possible. If N3 represents a Galactic micro-blazar, the increase in the brightness of the radio jet could account for the lack of X-ray emission observed in N3.


A micro-blazar origin for N3 predicts variability on short time scales due to large dependence of relativistic beaming on the particle velocity and beam orientation. Small changes in either would result in a large change in the observed intensity of N3. The micro-blazar hypothesis could be easily tested by examining N3 for short term radio variability.  

N3, despite our analysis, continues to be shrouded in mystery.  Although a micro-blazar hypothesis cannot be ruled out by our data, additional observations will be necessary to distinguish between a micro-blazar and the possibility that N3 represents an object not considered in this work. Future radio observations will be essential for uncovering additional clues that might lead to a viable working hypothesis.  For example, the character and timescale of the variability, as well as precise measurement of the spectral energy distribution will be valuable for taking the next steps. Furthermore, high-resolution VLBI observations may reveal structure in N3 because of its very compact size, although because of scatter-broadening by the foreground interstellar medium, such observations would need to be carried out at high radio frequencies.  Meanwhile, the N3 molecular cloud should also be observed to search for any weak atomic or molecular absorption of the continuum emission from N3, maser emission species other than \meth{}, or molecular line indicators that might reveal whether the molecular gas projected near N3 shows any excess heating.  

\section{Conclusions}

Using the VLA, we have conducted the first in-depth study of the GC radio source N3. Our observations reveal that:

(1) N3 is an extremely compact source and is unresolved at all frequencies in our data. Using our 49 GHz data, we place an upper limit of 0.0028$\times$0.0013 pc or 550$\times$250 AU (at a distance of 8.3 kpc) on the size of N3.

(2) The brightness of N3 is measured to be 62 mJy at 10.5 GHz, making N3 the brightest source in the radio arc region at these frequencies. Furthermore, the brightness of N3 has increased by more than a factor of three since the observations of \citet{Yusef-Zadeh1987} and \citet{Lang1997}.

(3) The spectrum of N3 can be modeled as a broken power law peaking near 8.68 GHz.  Our fit to the spectrum of N3 is consistent with self-absorbed synchrotron emission. 

(4) Adjacent to N3 in projection lies a compact molecular cloud, in which we detect 13 molecular transitions. The cloud morphology is suggestive of an interaction with N3. The high temperatures (T$_{45}=165$ K) and large linewidths (20-30 \kms{}) strongly suggest that this molecular cloud is located in the GC.

(5) Weak molecular emission, and no absorption is detected along the line of sight to N3. Additionally, a weak 36 GHz collisionally excited methanol maser is found to be positionally coincident with the continuum emission from N3. Combined, these two observations suggest that N3 lies within the near side of the molecular cloud. The location of the molecular cloud within the GC then suggests that N3 also lies within the GC. 

(6) The ``arcs'' in the southern portion of a thermal ``wake'' structure seen in Pa-$\alpha$ images appear to extend northward from the position of N3 and may show evidence of a past energetic outburst from N3. 

(7) We are able to rule out a UC\hii{} region, young supernova, nearby active star, AGN, and micro-quasar as possible physical counterparts for N3. While a micro-blazar explanation cannot be completely ruled out, more observations are needed to discern between a micro-blazar model and more exotic or unknown phenomena not considered in this work.

\section{Acknowledgements}

We wish to thank the staff of the Karl G. Jansky Very Large Array. 
CCL would like to the acknowledge the support of a Career Development Award from the University of Iowa and the School of Maths and Physics at the University of Tasmania. 
We also thank Philip Kaaret for his advice in examining the Chandra X-ray data and discussing the possibility of a micro-quasar as the origin of N3.

\bibliographystyle{aasjournal} 
\bibliography{N3_references.bib}

\clearpage

\section{Figures and Tables}

\begin{table}[ht]
\caption{Observed Fields} 
\begin{tabular}{ccccccc}
\\[0.5ex]
\hline\hline
 
 {\bf Field} & {\bf RA } &{\bf Declination } & \multicolumn{4}{c}{\bf Integration Time\footnotemark[1]}  \\
 {\bf Name} & (J2000) & (J2000) &  DnC\footnotemark[2] & CnB\footnotemark[3] & B\footnotemark[4] & BnA\footnotemark[5] \\
\hline
S-1 & 17$^{\mathrm{h}}$46$^{\mathrm{m}}$32.00$^{\mathrm{s}}$  & $-28\degr 51\arcmin 16.0\arcsec$ & 10.1 & 47.2 &  & 13.5  \\
\hline
C-1 & 17$^{\mathrm{h}}$46$^{\mathrm{m}}$35.92$^{\mathrm{s}}$  & $-28\degr 52\arcmin 05.3\arcsec$ & 10.1 & 11.9 & 29.9 & 11.5 \\
C-2 & 17$^{\mathrm{h}}$46$^{\mathrm{m}}$21.96$^{\mathrm{s}}$  & $-28\degr 49\arcmin 55.4\arcsec$ & 10.4 & 12.3 & 29.7 & 11.3 \\
\hline
X-1 & 17$^{\mathrm{h}}$46$^{\mathrm{m}}$41.79$^{\mathrm{s}}$  & $-28\degr 53\arcmin 21.9\arcsec$ & 10.2 & 9.9 & 32.7 & \\
X-2 & 17$^{\mathrm{h}}$46$^{\mathrm{m}}$32.63$^{\mathrm{s}}$  & $-28\degr 51\arcmin 49.6\arcsec$ & 10.4 & 10.3 & 32.3 & \\
X-3 & 17$^{\mathrm{h}}$46$^{\mathrm{m}}$24.21$^{\mathrm{s}}$  & $-28\degr 50\arcmin 13.6\arcsec$ & 10.4 & 10.3 & 32.3 & \\
X-4 & 17$^{\mathrm{h}}$46$^{\mathrm{m}}$15.20$^{\mathrm{s}}$  & $-28\degr 48\arcmin 45.5\arcsec$ & 10.4 & 10.3 & 32.3 & \\
\hline

Ka-1  & 17$^{\mathrm{h}}$46$^{\mathrm{m}}$24.34$^{\mathrm{s}}$  & $-28\degr 48\arcmin 57.1\arcsec$ & 9.6 &  &  & \\ 
Ka-2  & 17$^{\mathrm{h}}$46$^{\mathrm{m}}$19.61$^{\mathrm{s}}$  & $-28\degr 50\arcmin 31.8\arcsec$ & 9.9 &  &  & \\  
Ka-3  & 17$^{\mathrm{h}}$46$^{\mathrm{m}}$21.96$^{\mathrm{s}}$  & $-28\degr 49\arcmin 44.8\arcsec$ & 9.9 &  & 40.3 & \\ 

\hline

Q-1  & 17$^{\mathrm{h}}$46$^{\mathrm{m}}$19.12$^{\mathrm{s}}$  & $-28\degr 50\arcmin 40.7\arcsec$ & 14.3 &  &  &  \\ 
Q-2  & 17$^{\mathrm{h}}$46$^{\mathrm{m}}$21.02$^{\mathrm{s}}$  & $-28\degr 50\arcmin 03.2\arcsec$ & 14.3 &  & 56.3 \\ 
Q-3  & 17$^{\mathrm{h}}$46$^{\mathrm{m}}$22.90$^{\mathrm{s}}$  & $-28\degr 49\arcmin 25.9\arcsec$ & 14.3 &  &  & \\ 
Q-4  & 17$^{\mathrm{h}}$46$^{\mathrm{m}}$24.76$^{\mathrm{s}}$  & $-28\degr 48\arcmin 47.9\arcsec$ & 14.3 &  &  & \\ 

\hline

\end{tabular}
\footnotetext[1]{Integration time in minutes}
\footnotetext[2]{Observations made in May and June 2013.}
\footnotetext[3]{Observations made in September 2013.}
\footnotetext[4]{Observations made in November 2013.}
\footnotetext[5]{Observations made in February 2014.}
\label{pointings}
\end{table}

\begin{table}[hb]
\caption{Spectral Line Parameters} 
\begin{tabular}{cccccccc}
\\[0.5ex]
\hline\hline
{\bf Species+}  & {\bf Rest } & {\bf Channel} &{\bf Detection } \\
{\bf Transition} & {\bf Frequency} & {\bf Width}&  \\
& (GHz) &  (\kms) &  \\
\hline
H56-$\alpha$ & 36.4663 & 2.06 & No\\
H60-$\alpha$ & 30.7004 & 5.05 & No\\
SO (1$_0$--0$_1$) & 30.0016 & 5.00 & Yes     \\
\methts{} &  36.1693 & 0.26 &Yes  \\   
\cyano{}  (4--3) &  36.3923 & 2.06 &Yes  \\
\methcy{} (2$_0$--1$_0$)  &  36.7955 & 2.037 &Yes\\     
SiO (1--0) & 43.4238 & 2.25 & Yes\\
HNCO (2$_0$--1$_0$) & 43.9630 & 3.34 & Yes\\
\form{} ($4_{13}-4_{14}$) & 48.2846 & 3.10 & No\\
CS (1--0) & 48.9909 & 1.51 & Yes\\
\meth{} (1$_0$--0$_0$) & 48.3725 & 3.04 &Yes\\
\methff{} & 44.0694 & 0.21 & Yes \\
\hline\hline
\end{tabular}
\label{spec_image}
\end{table}
\clearpage

\begin{table}[ht]
\caption{36 GHz \meth{} Masers}
\centering
\begin{tabular}{cccccccccc}
\\[0.5ex]
	
\hline\hline
& & & & & & & & & \\
\multirow{2}{*}{\bf ID} & \multirow{2}{*}{\bf Maser     Name}&  \multirow{2}{*}{\bf RA}&        \multirow{2}{*}{\bf Dec}&       \multirow{2}{*}{\bf Velocity}&  \multirow{2}{*}{\bf FWHM}&      \multirow{2}{*}{\bf I$_{peak}$}&         \multirow{2}{*}{\bf Flux}&     \multirow{2}{*}{\bf Tb}&        {\bf 44 GHz}\\
& & & & & & & & & {\bf Counterpart}\\
& & HH:MM:SS.s	 & DD:MM:SS.s & \kms{} & 	\kms{}	 & \jyb{} & 	Jy \kms{} & K	 & \\
\hline
M36-1&    G0.1699558-0.0815683   &17:46:20.558   &-28:50:00.56   & 12.428   & 1.624   &  0.440   &  1.176&     7060   &  \\
M36-2&    G0.1696162-0.0812026   &17:46:20.424   &-28:50:00.92   & -6.221   & 1.529   &  0.354   &  1.614&     5675   &  \\
M36-3&    G0.1723175-0.0813672   &17:46:20.847   &-28:49:52.92   & -6.221   & 0.745   &  0.286   &  0.513&     4592   &  \\
M36-4&    G0.1695547-0.0846931   &17:46:21.232   &-28:50:07.64   & 24.961   & 1.361   &  0.151   &  0.541&     2418   & M44-2 \\
M36-5&    G0.1666236-0.0823013   &17:46:20.255   &-28:50:12.18   & -2.177   & 1.270   &  0.142   &  0.278&     2280   &  \\
M36-6&    G0.1666081-0.0822652   &17:46:20.245   &-28:50:12.16   & -1.041   & 1.519   &  0.128   &  0.244&     2049   &  \\
M36-7&    G0.1686226-0.0798366   &17:46:19.963   &-28:50:01.42   &  5.276   & 1.125   &  0.105   &  0.115&     1688   &  \\
M36-8&    G0.1691074-0.0830598   &17:46:20.786   &-28:50:05.96   & 15.537   & 1.136   &  0.101   &  0.165&     1619   & M44-11 \\
M36-9&    G0.1706044-0.0851145   &17:46:21.481   &-28:50:05.20   &-13.056   & 0.831   &  0.100   &  0.185&     1607   &  \\
M36-10&   G0.1686274-0.0798337   &17:46:19.963   &-28:50:01.40   &  4.140   & 1.213   &  0.090   &  0.111&     1448   &  \\
M36-11&   G0.1699529-0.0815635   &17:46:20.557   &-28:50:00.56   &  8.384   & 0.788   &  0.089   &  0.168&     1423   &  \\
M36-12&   G0.1696997-0.0826865   &17:46:20.783   &-28:50:03.44   &  3.104   & 0.905   &  0.069   &  0.145&     1113   &  \\
M36-13&   G0.1690756-0.0833068   &17:46:20.840   &-28:50:06.52   & 11.910   & 1.275   &  0.067   &  0.073&     1079   &  \\
M36-14&   G0.1693958-0.0848808   &17:46:21.254   &-28:50:08.48   &-13.056   & 1.591   &  0.065   &  0.121&     1049   & M44-12 \\
M36-15&   G0.1701438-0.0814604   &17:46:20.560   &-28:49:59.78   & 11.492   & $<$0.518   &  0.059   &  0.028&      946   &  \\
M36-16&   G0.1702570-0.0834530   &17:46:21.042   &-28:50:03.16   & 10.974   & 1.343   &  0.059   &  0.092&      938   &  \\
\hline
\end{tabular}
\label{36Masers}
\end{table}

\begin{table}[hb]
\caption{44 GHz \meth{} Masers}
\centering
\begin{tabular}{cccccccccc}
\\[0.5ex]
\hline\hline
& & & & & & & & & \\
\multirow{2}{*}{\bf ID} & \multirow{2}{*}{\bf Maser	Name}&	\multirow{2}{*}{\bf RA}&	\multirow{2}{*}{\bf Dec}&	\multirow{2}{*}{\bf Velocity}&	\multirow{2}{*}{\bf FWHM}&	\multirow{2}{*}{\bf I$_{peak}$}&	 \multirow{2}{*}{\bf Flux}&	\multirow{2}{*}{\bf Tb}&	{\bf 36 GHz}\\
& & & & & & & & & {\bf Counterpart}\\
& & HH:MM:SS.s	 & DD:MM:SS.s & 	\kms{} & 	\kms{}	 & \jyb{} & 	Jy \kms{} & K	 & \\
\hline
M44-1&   G0.1695479-0.0846845    &17:46:21.229   &-28:50:07.65   & 23.500   & 0.780   &  0.369   &  0.725&     6026   &  \\
M44-2&   G0.1695529-0.0846928    &17:46:21.232   &-28:50:07.65   & 25.000   & 0.954   &  0.354   &  0.819&     5773   & M36-4 \\
M44-3&   G0.1690229-0.0830920    &17:46:20.782   &-28:50:06.28   & 14.000   & 0.855   &  0.327   &  0.702&     5337   &  \\
M44-4&   G0.1719719-0.0810931    &17:46:20.734   &-28:49:53.47   &  7.000   & 1.668   &  0.180   &  0.648&     2933   &  \\
M44-5&   G0.1688294-0.0830731    &17:46:20.750   &-28:50:06.84   & 18.000   & 1.013   &  0.145   &  0.357&     2363   &  \\
M44-6&   G0.1705161-0.0810569    &17:46:20.518   &-28:49:57.88   &  3.000   & 0.566   &  0.111   &  0.140&     1808   &  \\
M44-7&   G0.1688294-0.0830731    &17:46:20.750   &-28:50:06.84   & 19.500   & $<$ 0.500   &  0.109   &  0.100&     1774   &  \\
M44-8&   G0.1700713-0.0822152    &17:46:20.726   &-28:50:01.42   &  1.500   & 0.708   &  0.108   &  0.156&     1764   &  \\
M44-9&   G0.1719770-0.0811014    &17:46:20.737   &-28:49:53.47   &  4.500   & 1.590   &  0.107   &  0.316&     1751   &  \\
M44-10&   G0.1700746-0.0822018   &17:46:20.723   &-28:50:01.38   &  3.500   & 1.334   &  0.101   &  0.106&     1655   & \\
M44-11&   G0.1691161-0.0830580   &17:46:20.787   &-28:50:05.93   & 15.500   & 1.130   &  0.089   &  0.085&     1446   & M36-8 \\
M44-12&   G0.1693973-0.0848672   &17:46:21.251   &-28:50:08.45   &-12.500   & 1.367   &  0.076   &  0.115&     1248   & M36-14 \\
\hline
\hline
\end{tabular}
\label{44Masers}
\end{table}

\end{document}